\documentclass[twocolumn,aps,prl,showpacs]{revtex4}
\usepackage[T1]{fontenc}
\usepackage[latin9]{inputenc}
\usepackage{amsmath}
\usepackage{amssymb}
\usepackage{graphicx}
\usepackage{esint}

\makeatletter
\@ifundefined{textcolor}{}
{%
 \definecolor{BLACK}{gray}{0}
 \definecolor{WHITE}{gray}{1}
 \definecolor{RED}{rgb}{1,0,0}
 \definecolor{GREEN}{rgb}{0,1,0}
 \definecolor{BLUE}{rgb}{0,0,1}
 \definecolor{CYAN}{cmyk}{1,0,0,0}
 \definecolor{MAGENTA}{cmyk}{0,1,0,0}
 \definecolor{YELLOW}{cmyk}{0,0,1,0}
}

\makeatother

\makeatother

\begin{document}

\title{Topological Fulde-Ferrell superfluid in spin-orbit coupled atomic
Fermi gases}

\author{Xia-Ji Liu$^{1}$}

\email{xiajiliu@swin.edu.au}

\author{Hui Hu$^{1}$}

\affiliation{$^{1}$ARC Centres of Excellence for Quantum-Atom Optics and Centre
for Atom Optics and Ultrafast Spectroscopy, Swinburne University of
Technology, Melbourne 3122, Australia}

\date{\today}
\begin{abstract}
We theoretically predict a new topological matter - topological inhomogeneous
Fulde-Ferrell superfluid - in one-dimensional atomic Fermi gases with
equal Rashba and Dresselhaus spin-orbit coupling near $s$-wave Feshbach
resonances. The realization of such a spin-orbit coupled Fermi system
has already been demonstrated recently by using a two-photon Raman
process and the extra one-dimensional confinement is easy to achieve
using a tight two-dimensional optical lattice. The topological Fulde-Ferrell
superfluid phase is characterized by a nonzero center-of-mass momentum
and a non-trivial Berry phase. By tuning the Rabi frequency and the
detuning of Raman laser beams, we show that such an exotic topological
phase occupies a significant part of parameter space and therefore
it could be easily observed experimentally, by using, for example,
momentum-resolved and spatially resolved radio-frequency spectroscopy.
\end{abstract}


\maketitle
Topological superfluids attract tremendous interests over the past
few years \cite{Qi2011}. In addition to providing a new quantum phase
of matter, topological superfluids can host exotic quasiparticles
at their boundary, which are known as Majorana fermions - particles
that are their own antiparticles \cite{Majorana1937,Wilczek2009}.
Due to their non-Abelian exchange statistics, Majorana fermions are
believed to be the essential quantum bits for topological quantum
computation \cite{Nayak2008}. Therefore, the pursuit for topological
superfluids and Majorana fermions turns out to be one of the most
important challenges in fundamental science. Theoretically, a number
of settings have been proposed for the realization of topological
superfluids, including the fractional quantum Hall states at filling
$\nu=5/2$ \cite{Moore1991}, vortex states of $p_{x}+ip_{y}$ superconductors
\cite{Read2000,Mizushima2008}, and surfaces of three-dimensional
(3D) topological insulators in proximity to an $s$-wave superconductor
\cite{Sau2010}, and one-dimensional (1D) nanowires with strong spin-orbit
coupling coated also on an $s$-wave superconductor \cite{Orge2010}.
In the latter setting, in-direct evidences of topological superfluid
and Majorana fermions have been observed experimentally \cite{Mourik2012}.

Ultracold Fermi gas with spin-orbit coupling near an $s$-wave Feshbach
resonance is a new promising candidate to create topological superfluids
\cite{Zhang2008,Liu2012a,Liu2012b,Liu2013a,Wei2012}. Due to the unprecedented
controllability in interatomic interaction, dimensionality and purity,
there are a number of rapid experimental advances \cite{Bloch2008}.
In particular, a spin-orbit coupled Fermi gas can now be routinely
realized by using two counterpropagating Raman laser beams \cite{Wang2012,Cheuk2012,Willianms2013},
a scheme first advanced by Ian Spielman and co-workers \cite{Lin2011}.
Through the use of Feshbach resonances \cite{Fu2013} and optical
lattices \cite{Liao2010}, a strongly interacting spin-orbit coupled
Fermi gas in low dimensions could be manipulated immediately. Theoretical
proposals for engineering a topological superfluid in such Fermi gas
systems have been discussed in greater detail by a number of cold-atom
researchers \cite{Zhang2008,Liu2012a,Liu2012b,Liu2013a,Wei2012}.

\begin{figure}[t]
\begin{centering}
\includegraphics[clip,width=0.45\textwidth]{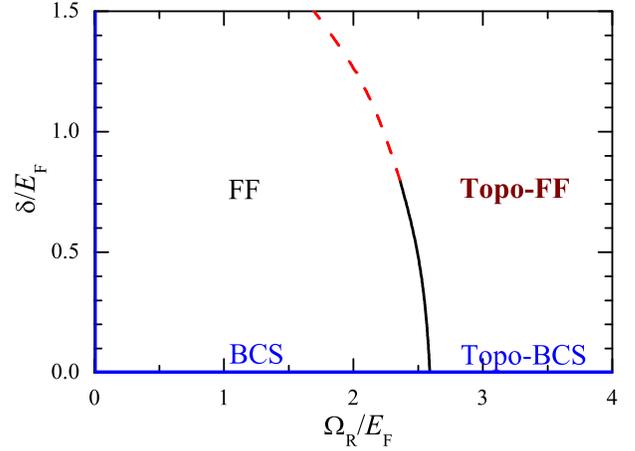} 
\par\end{centering}

\caption{(color online) Zero-temperature phase diagram of a 1D spin-orbit coupled
atomic Fermi gas near a broad Feshbach resonance. In the presence
of a synthetic spin-orbit coupling induced by two counter-propagating
Raman laser beams, a topologically non-trivial Fulde-Ferrell superfluid
appears when the Raman Rabi frequency $\Omega_{R}$ is above a threshold
at finite detunings $\delta$. Here we take the recoil momentum $k_{R}=1.25k_{F}$
and a dimensionless interaction parameter $\gamma=3$. Depending on
the detuning, the transition could be either continuous (solid line)
or of first order (dashed line). The FF superfluid reduces to a BCS
superfluid when $\Omega_{R}=0$ or $\delta=0$.}

\label{fig1} 
\end{figure}

In this work, we theoretically predict a new type topological superfluid,
in which the superfluid order parameter varies in real space \cite{Fulde1964,Larkin1964}.
This prediction is motivated by the recent discovery that by imposing
an in-plane Zeeman field along one of the directions of synthetic
spin-orbit coupling, the phase space for inhomogeneous Fulde-Ferrell
(FF) superfluidity is greatly enlarged \cite{Zheng2013,Wu2013,Liu2013b,Dong2013}.
In low dimensions, this inhomogeneous superfluid may acquire non-trivial
topological feature. Our main result is summarized in Fig. \ref{fig1},
which shows a zero-temperature phase diagram for a 1D interacting
Fermi gas with the experimentally realized equal Rashba and Dresselhaus
spin-orbit coupling \cite{Wang2012,Cheuk2012}. We find a large window
for the topological inhomogeneous FF superfluid, characterized by
both a nonzero center-of-mass momentum and a non-trivial Berry phase.
This exotic superfluid phase could be easily created and probed in
current experiments, once the heating issue related to the Raman process
is overcome. We may also anticipate the appearance of topological
inhomogeneous superfluid in two dimensions, but with Rashba spin-orbit
coupling, which is so far not experimentally realized yet.

Our investigation is based on the mean-field theory which is qualitatively
reliable at zero temperature. It does not catpure the large phase
fluctuations found in 1D, although the mean-field physics is robust
against these fluctuations as shown by Ref. \cite{Fidkowski2011}.
More accurate description could be obtained by using other standard
techniques in 1D, for example, bosonization. In that language, the
ground state may be identified as a two-component Luttinger liquid
\cite{Zhao2008}. These possibilities will be addressed in the future
study.

\textit{Model Hamiltonian and mean-field theory.} --- We consider
a 1D spin-orbit coupled two-component Fermi gas near a broad Feshbach
resonance. Experimentally, the 1D confinement can be easily created
by imposing a tight 2D optical lattice \cite{Liao2010}, i.e., in
the $x-y$ plane. The synthetic spin-orbit coupling has already been
engineered in $^{6}$Li or $^{40}$K atoms by using the Raman scheme
first demonstrated at NIST for a $^{87}$Rb Bose-Einstein condensate
(BEC) \cite{Lin2011}. In this scheme, two Raman laser beams counter-propagate
along the $z$-direction and couple the two different hyperfine states,
giving rise to the term in the model Hamiltonian, $(\Omega_{R}/2)\int d{\bf x[}\Psi_{\uparrow}^{\dagger}(z)e^{i2k_{R}z}\Psi_{\downarrow}(z)+$H.c.$]$,
where $\Psi_{\sigma}^{\dagger}(z)$ is the creation field operator
for atoms in one of the hyperfine states $\sigma$ ($=\left|\downarrow\right\rangle ,\left|\uparrow\right\rangle $,
referred to as the spin state), $\Omega_{R}$ is the Rabi frequency
of Raman beams, and $k_{R}$ $=2\pi/\lambda_{R}$ is the recoil momentum
determined by the wave length $\lambda_{R}$ of two beams. Thus, during
the two-photon Raman process, atoms absorb a momentum of $2\hbar k_{R}$
and simultaneously change their spin state from $\left|\downarrow\right\rangle $
to $\left|\uparrow\right\rangle $, creating a correlation between
spin and orbital motion. This can be seen most clearly by introducing
a gauge transformation, $\Psi_{\uparrow}(z)=e^{ik_{R}z}\psi_{\uparrow}(z)$
and $\Psi_{\downarrow}(z)=e^{-ik_{R}z}\psi_{\downarrow}(z)$, which
leads to a term proportional to $\hat{k}\sigma_{z}$, where $\hat{k}\equiv-i\partial_{z}$
and $\sigma_{z}$ is the Pauli matrix. Near a broad Feshbach resonance,
the spin-orbit coupled Fermi gas system may therefore be described
by a single-channel model Hamiltonian ${\cal H}=\int dz[{\cal H}_{0}+{\cal H}_{int}]$,
where the single-particle part 
\begin{equation}
{\cal H}_{0}=\left[\psi_{\uparrow}^{\dagger},\psi_{\downarrow}^{\dagger}\right]\left[\begin{array}{cc}
\hat{\xi}_{k}+\lambda\hat{k}+\delta/2 & \Omega_{R}/2\\
\Omega_{R}/2 & \hat{\xi}_{k}-\lambda\hat{k}-\delta/2
\end{array}\right]\left[\begin{array}{c}
\psi_{\uparrow}\\
\psi_{\downarrow}
\end{array}\right],\label{eq:spHami}
\end{equation}
and ${\cal H}_{int}=g_{1D}\psi_{\uparrow}^{\dagger}(z)\psi_{\downarrow}^{\dagger}(z)\psi_{\downarrow}(z)\psi_{\uparrow}(z)$
is the interaction Hamiltonian that describes the contact interaction
between two spin states with 1D effective interaction strength $g_{1D}<0$
\cite{Bergeman2003}. 

In the single-particle Hamiltonian (\ref{eq:spHami}), $\hat{\xi_{k}}\equiv-\hbar^{2}\partial_{z}^{2}/(2m)-\mu$
is the kinetic energy after we drop a constant recoil energy $E_{R}\equiv\hbar^{2}k_{R}^{2}/(2m)$,
and $\delta$ is the two-photon detuning from the Raman resonance.
For convenience, we have defined a spin-orbit coupling constant $\lambda\equiv\hbar^{2}k_{R}/m$.
The strength of spin-orbit coupling may be characterized by a dimensionless
coupling constant $\tilde{\lambda}=\lambda k_{F}/E_{F}$, where $k_{F}$
is the Fermi wave-vector and $E_{F}\equiv\hbar^{2}k_{F}^{2}/(2m)$
is the Fermi energy. In the Shanxi experiment with $^{40}$K atoms
\cite{Wang2012}, the Fermi wavelength is typically about $k_{F}\simeq k_{R}$
and the Rabi frequency $\Omega_{R}\simeq2E_{R}$. In the quasi-1D
geometry formed by a tight 2D optical lattice \cite{Liao2010}, it
is shown by Olshanii and co-workers that the effective interaction
strength $g_{1D}$ could be expressed through a 3D $s$-wave scattering
length $a_{3D}$ \cite{Bergeman2003}. It is useful to characterize
the interaction strength $g_{1D}$ by using a dimensionless interaction
parameter $\gamma\equiv-mg_{1D}/(\hbar^{2}n)$, where $n=2k_{F}/\pi$
is the linear density in 1D.

In the absence of detuning ($\delta=0$), the model Hamiltonian (\ref{eq:spHami})
for cold-atoms has been previously solved by the present authors \cite{Liu2012b,Liu2013a}
and Mueller and co-workers \cite{Wei2012}. It is known that when
the Rabi frequency $\Omega_{R}$ is above a threshold $(\Omega_{R}/2)_{c}=\sqrt{\Delta^{2}+\mu^{2}}$,
where $\Delta$ is the pairing gap, the system becomes a topological
superfluid. The 3D counterpart of the model Hamiltonian has also been
investigated \cite{Liu2013b}. In this case, a nonzero detuning leads
to the inhomogeneous FF superfluid state in which Cooper pairs carry
a single valued center-of-mass momentum \cite{Fulde1964}. Therefore,
it is natural to anticipate that in 1D a topological inhomogeneous
FF superfluid may arise at finite detuning $\delta\neq0$ and at a
large Rabi frequency. In the following, we confirm this anticipation
by detailed numerical calculations.

For this purpose, we assume a FF-like order parameter $\Delta(z)=-g_{1D}\left\langle \psi_{\downarrow}(z)\psi_{\uparrow}(z)\right\rangle =\Delta e^{iqz}$
and consider the mean-field decoupling of the interaction Hamiltonian,
${\cal H}_{int}\simeq-[\Delta(z)\psi_{\uparrow}^{\dagger}(z)\psi_{\downarrow}^{\dagger}(z)+\textrm{H.c.}]-\Delta^{2}/g_{1D}$.
By using a Nambu spinor $\Phi(z)\equiv[\psi_{\uparrow}(z),\psi_{\downarrow}(z),\psi_{\uparrow}^{\dagger}(z),\psi_{\downarrow}^{\dagger}(z)]$$^{T}$,
the total Hamiltonian can be written into a compact form, $\mathcal{H}=(1/2)\int d{\bf x}\Phi^{\dagger}(z)\mathcal{H}_{BdG}\Phi(z)-L\Delta^{2}/U_{0}+\sum_{k}\hat{\xi}_{k}$,
where $L$ is the length of the system and the Bogoliubov Hamiltonian
takes the form 
\begin{equation}
\mathcal{H}_{BdG}\equiv\left[\begin{array}{cccc}
\mathcal{S}_{k}^{+} & \Omega_{R}/2 & 0 & -\Delta\left(z\right)\\
\Omega_{R}/2 & \mathcal{S}_{k}^{-} & \Delta\left(z\right) & 0\\
0 & \Delta^{*}\left(z\right) & -\mathcal{S}_{-k}^{+} & -\Omega_{R}/2\\
-\Delta^{*}\left(z\right) & 0 & -\Omega_{R}/2 & -\mathcal{S}_{-k}^{-}
\end{array}\right]\label{eq:BdGHami}
\end{equation}
with $\mathcal{S}_{k}^{\pm}\equiv\hat{\xi}_{k}\pm\lambda\hat{k}\pm\delta/2$.
It is straightforward to diagonalize the Bogoliubov Hamiltonian $\mathcal{H}_{BdG}\Phi_{k\eta}(z)=E_{k\eta}\Phi_{k\eta}(z)$
with quasiparticle wave-function $\Phi_{k\eta}(z)=e^{ikz}/\sqrt{L}[u_{k\eta\uparrow}e^{iqz/2},u_{k\eta\downarrow}e^{iqz/2},v_{k\eta\uparrow}e^{-iqz/2},v_{k\eta\downarrow}e^{-iqz/2}]^{T}$
and quasiparticle energy $E_{k\eta}$ ($\eta=1,2,3,4$). The mean-field
thermodynamic potential $\Omega$ at temperature $T$ is given by,
\begin{eqnarray}
\frac{\Omega}{L} & = & \frac{1}{2L}\left[\sum_{k}\left(\xi_{k+q/2}+\xi_{k-q/2}\right)-\sum_{k\eta}E_{k\eta}\right]\nonumber \\
 &  & -\frac{k_{B}T}{L}\sum_{k\eta}\ln\left(1+e^{-E_{k\eta}/k_{B}T}\right)-\frac{\Delta^{2}}{g_{1D}}.\label{eq:Omega}
\end{eqnarray}
Here $\xi_{k}\equiv\hbar^{2}k^{2}/(2m)-\mu$ and the summation over
quasiparticle energy should be restricted to $E_{k\eta}\geq0$ because
of an inherent particle-hole symmetry in the Nambu spinor representation
\cite{ParticleHoleSymmetry}. For a given set of parameters (i.e,
the temperature $T$, interaction strength $\gamma$ etc.), different
mean-field phases can be determined using the self-consistent stationary
conditions: $\partial\Omega/\partial\Delta=0$, $\partial\Omega/\partial q=0$,
as well as the conservation of total atom number, $N=nL=-\partial\Omega/\partial\mu$.
At finite temperatures, the ground state has the lowest free energy
$F=\Omega+\mu N$.

\textit{Topological Fulde-Ferrell superfluid.} --- Let us focus on
the phase diagram at zero temperature. The Raman Rabi frequency $\Omega_{R}$
and the detuning $\delta$ can be experimentally tuned \cite{Wang2012,Cheuk2012}.
Thus, we shall present the phase diagram as functions of $\Omega_{R}$
and $\delta$. Throughout the paper, we consider a realistic dimensionless
spin-orbit coupling constant. According to the typical number of atoms
in experiments \cite{Wang2012}, we take $k_{R}=5k_{F}/4$, corresponding
to a dimensionless spin-orbit coupling constant $\tilde{\lambda}=\lambda k_{F}/E_{F}=2.5$.
We also choose a typical 3D $s$-wave scattering length near the broad
Feshbach resonance and obtain the 1D effective interaction strength
by using the standard relation for confinement induced resonance \cite{Bergeman2003}.
This leads to a dimensionless interaction parameter $\gamma=3$ \cite{Liao2010,Liu2007,Liu2008}.

\begin{figure}[t]
\begin{centering}
\includegraphics[clip,width=0.45\textwidth]{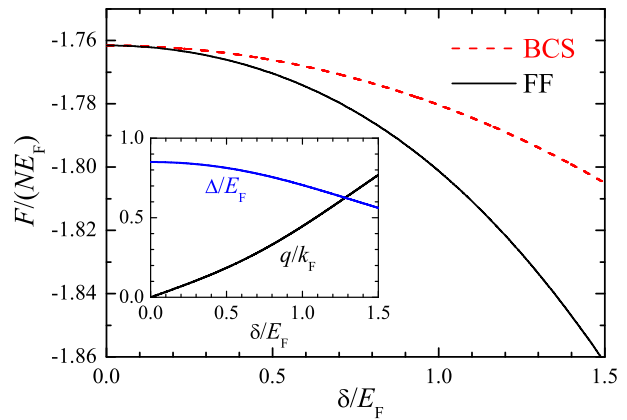} 
\par\end{centering}

\caption{(color online) Free energy of the FF superfluid (solid line) and of
the BCS superfluid (dashed line), as a function of detuning at a large
Rabi frequency$\Omega_{R}=3E_{F}$ and at $T=0$. To obtain the free
energy of the BCS superfluid, we have forced the center-of-mass momentum
to be strictly zero. The FF superfluid appears always to be energetically
favorable. The inset shows the detuning dependence of the pairing
gap and momentum of the FF superfluid state.}

\label{fig2} 
\end{figure}

It is known from the previous studies that, in a 3D Fermi gas with
equal Rashba and Dresselhaus spin-orbit coupling, a FF superfluid
with a single-valued center-of-mass momentum (i.e., the FF momentum)
is aways energetically favorable at any finite detuning and nonzero
Rabi frequency \cite{Liu2013b}. This superfluid corresponds to the
solution with $\Delta\neq0$ and $q\neq0$. In Fig. \ref{fig2}, we
check that this observation also holds in 1D, where the transverse
spatial degree of freedom of the atoms is frozen by a tight 2D optical
lattice. Fig. \ref{fig2} compares the free energy of a FF superfluid
and of a standard BCS superfluid at the Rabi frequency $\Omega_{R}=3E_{F}$.
The solution of the BCS superfluid is obtained by artificially restricting
the FF momentum $q=0$. A normal state is also considered, but is
found to have higher energy than superfluid phases and is only favorable
at sufficiently large detuning and/or Rabi frequency. All the three
competing phases (normal, BCS or FF) are stable against phase separation
(i.e., $\partial^{2}\Omega/\partial\Delta^{2}\geq0$). Thus, we do
not include the phase-separation phase which involves two or more
competing phases. It is clear from Fig. \ref{fig2} that the free
energy of FF superfluid is lower by an amount of $0.05NE_{F}$ than
that of the BCS superfluid at the typical detuning $\delta=1.5E_{F}$,
suggesting that the FF superfluid should also be energetically favorable
at finite temperatures below the superfluid phase transition which
would occur at about one-tenth of the Fermi temperature \cite{Liu2013a}.
The detuning dependence of the FF pairing gap and momentum is shown
in the inset of Fig. \ref{fig2}. The FF momentum increases rapidly
with increasing the detuning. At the same time, the pairing gap decreases
as the detuning behaves like an effective Zeeman field. The system
will become a normal Fermi gas at sufficiently large detuning beyond
the so-called Chandrasekhar-Clogston limit.

Now, let us investigate in greater detail the properties of the FF
superfluid. At zero detuning, where $q=0$, the FF superfluid reduces
to a BCS superfluid \cite{Liu2013b}. In this limit, it is known that
there is a topological phase transition at a threshold Rabi frequency
$(\Omega_{R})_{c}$ \cite{Liu2012b,Liu2013a,Wei2012}. This threshold
corresponds to a critical point where the energy gap close and then
open. The topology of the Fermi surface of the system changes beyond
the critical point, by developing non-trivial spin texture in real
space characterized by Berry phase or topological invariant. Indeed,
by analyzing the analytical expression of the energy spectrum at $\delta=0$,
it is easy to obtain $(\Omega_{R})_{c}/2=\sqrt{\Delta^{2}+\mu^{2}}$.
By increasing the detuning of Raman beams, the BCS superfluid changes
smoothly into a FF superfluid with a nonzero FF momentum, $q\neq0$
\cite{Liu2013b}. In this case, it is reasonable to argue that the
procedure of closing and opening energy gap may persist. Thus, the
topology of the FF superfluid will change accordingly. As a result,
we may obtain a topological inhomogeneous FF superfluid. 

\begin{figure}[t]

\begin{centering}
\includegraphics[clip,width=0.45\textwidth]{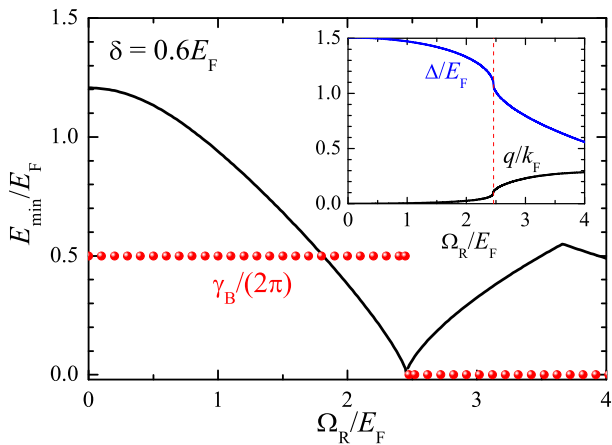} 
\par\end{centering}

\caption{(color online) Theoretical examination of the topological phase transition
at the detuning $\delta=0.6E_{F}$. The transition occurs at $\Omega_{R}\simeq2.46E_{F}$
, where the energy gap of the system (solid line) close and then open.
The Berry phase $\gamma_{B}$ is $\pi$ and 0 at the topologically
trivial and non-trivial regimes (circles). The insets shows the order
parameter and momentum of the FF superfluid, as a function of the
Rabi frequency.}

\label{fig3} 
\end{figure}

\begin{figure}[t]
\begin{centering}
\includegraphics[clip,width=0.45\textwidth]{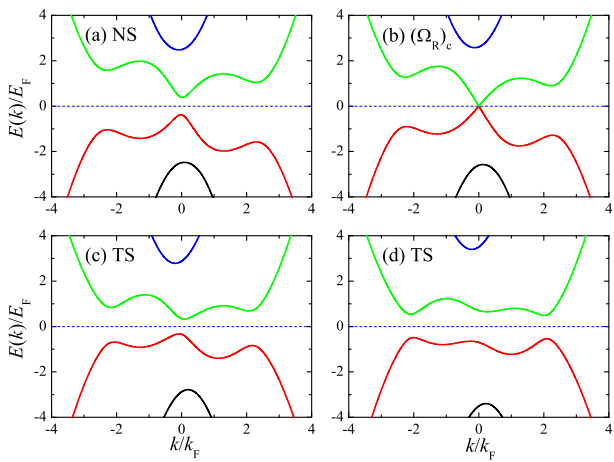} 
\par\end{centering}

\caption{(color online) The excitation spectrum $E_{i}(k)$ ($i=1,2,3$ and
$4$, from bottom to top) at different Rabi frequencies: (a) $\Omega_{R}=2.0E_{F}$,
(b) the critical field $\Omega_{R}\simeq2.45E$, (c) $\Omega_{R}=3.0E_{F}$
and (d) $\Omega_{R}=4.0E_{F}$. Here we take $\delta=0.6E_{F}$. Due
to the particle-hole symmetry, $E_{2}(k)=-E_{3}(-k)$ and $E_{1}(k)=-E_{4}(-k)$. }

\label{fig4} 
\end{figure}

In Fig. \ref{fig3}, we present the energy gap of the FF superfluid
at a nonzero detuning $\delta=0.6E_{F}$ as a function of the Rabi
frequency. With increasing the Rabi frequency, the energy gap close
and then open at the threshold $(\Omega_{R})_{c}\simeq2.45E_{F}$.
The four excitation spectra of Bogoliubov quasiparticles are shown
in Fig. \ref{fig4} at different Rabi frequencies. At the threshold,
as shown in Fig. \ref{fig4}(b), the bottom (top) of the second (third)
spectrum of hole (particle) excitations touches zero, giving rise
to a zero energy gap. Before and after this threshold, see for example,
Fig. \ref{fig4}(a) and \ref{fig4}(c) respectively, the excitation
spectrum is more or less the same. Both of them are gapped. The asymmetry
of the excitation spectrum with respect to the point $k=0$ is due
to the nonzero detuning and FF momentum. In the inset of Fig. \ref{fig3},
we also show the pairing gap and FF momentum with increasing the Rabi
frequency. Across the topological transition point, the pairing gap
and FF momentum decreases and increases more rapidly as the Rabi frequency
increases.

To characterize in a more quantitative way the topological phase transition,
we calculate the Berry phase defined by \cite{Wei2012} 
\begin{equation}
\gamma_{B}=i\intop_{-\infty}^{+\infty}dk\left[W_{1}^{*}(k)\partial_{k}W_{1}(k)+W_{2}^{*}(k)\partial_{k}W_{2}(k)\right].\label{eq:BerryPhase}
\end{equation}
Here $W_{\eta}(k)$ is the wave function of the $\eta$-th energy
band: $W_{\eta}(k)\equiv[u_{k\eta\uparrow}e^{iqz/2},u_{k\eta\downarrow}e^{iqz/2},v_{k\eta\uparrow}e^{-iqz/2},v_{k\eta\downarrow}e^{-iqz/2}]^{T}$.
In Fig. \ref{fig3}, the Berry phase is shown by circles. It jumps
from $\pi$ to $0$, right across the topological superfluid transition.
It is somewhat counter-intuitive the $\gamma_{B}=0$ sector corresponds
to the topologically non-trivial superfluid state. But it agrees well
with the known result in the limit of zero detuning \cite{Wei2012},
where a standard BCS superfluid translates into a topological BCS
superfluid. 

\textit{Experimental detection.} --- To experimentally probe the topological
inhomogeneous superfluid, one should measure the nonzero FF momentum
$q$ and at the same time confirm the non-trivial topological feature
of the system. For the former, the nonzero FF momentum might be determined
by using momentum-resolved radio-frequency spectroscopy, with a similar
scenario for a 3D spin-orbit coupled atomic Fermi gas, where the value
of $q/2$ can be directly read from the spectroscopy \cite{Liu2013b}.
To confirm the non-trivial topological feature, we may consider using
a spatially resolved radio-frequency spectroscopy to image the resulting
Majorana fermions localized at the boundary of the Fermi gas system
\cite{Liu2012b,Liu2013a,Wei2012}. In reality, both the finite temperature
and the existence of a harmonic trap will decrease the signal in the
spectroscopy. These issues will be addressed in the later study by
solving the Bogoliubov de-Gennes equation of a harmonically trapped
1D spin-orbit coupled Fermi gas with nonzero laser detuning at finite
temperatures.

\textit{Conclusions.} --- In summary, we have proposed that a new
topological superfluid with inhomogeneous order parameter in real
space could be potentially realized in current cold-atom settings,
by using a one-dimensional atomic Fermi gas with equal Rashba and
Dresselhaus spin-orbit coupling that has already been created in laboratories
through two-photon Raman process \cite{Wang2012,Cheuk2012}. The one
dimensional confinement is straightforward to implement via a tight
two-dimensional optical lattice, as demonstrated recently by Randy
Hulet group at Rice University \cite{Liao2010}. We believe such a
topological inhomogeneous superfluid is within the reach in the near
future once the temperature of the Fermi cloud could be cooled down
to one-tenth of the Fermi temperature. The inhomogeneity and the non-trivial
topological feature of the superfluid can be revealed by using momentum-resolved
and spatially resolved radio-frequency spectroscopies, respectively.
It is also of great interest to predict similar topological inhomogeneous
superfluid in solid-state systems, but the unprecedented controllability
in cold-atom systems would pave a new way to explore the long-sought
topological and inhomogeneous superfluidity.

\textit{Acknowledgements.} --- We are grateful to Han Pu, Lin Dong
and Su Yi for stimulating discussions. This research was supported
by the ARC Discovery Projects (DP0984637, DP0984522) and the NFRP-China
2011CB921502.

\textit{Note added.} --- In completing this work, we are aware of
related works which theoretically propose the topological inhomogeneous
superfluid in a 1D or 2D spin-orbit coupled Fermi gas with additional
optical lattices \cite{Chen2013,Qu2013}, or in a 2D Rashba spin-orbit
coupled Fermi gas in free space \cite{Zhang2013}.


\begin{thebibliography}{10}
\bibitem{ParticleHoleSymmetry}It is straightforward to check that
for any particle solution $[u_{k\uparrow},u_{k\downarrow},v_{k\uparrow},v_{k\downarrow}]$$^{T}$
with energy $E_{k}\geq0$, there is a one-to-one corresponding hole
solution $[v_{-k\uparrow}^{*},v_{-k\downarrow}^{*},u_{-k\uparrow}^{*},u_{-k\downarrow}^{*}]$$^{T}$
with energy $-E_{-k}$. These two solutions correspond to the same
physical state.
\end{thebibliography}
\end{document}